\documentclass[]{aa} %
\usepackage{graphicx}
\usepackage{txfonts}
\usepackage{verbatim}
\usepackage{bm}

\newcommand{\figref}[1]{Fig.~\ref{#1}}

\begin{document}
   \title{NIKA: A millimeter-wave kinetic inductance camera}


   \author{
    A. Monfardini\inst{1}\fnmsep\thanks{e-mail: monfardini@grenoble.cnrs.fr}, L. J. Swenson\inst{1},
    A. Bideaud\inst{1}, F. X. D\'esert\inst{5}, S. J. C. Yates\inst{3}, A. Benoit\inst{1},
    A. M. Baryshev\inst{2}, J. J. A. Baselmans\inst{3}, S. Doyle\inst{6},
    B. Klein\inst{7}, M. Roesch\inst{8},  C. Tucker\inst{6},
    P. Ade\inst{6}, M. Calvo\inst{4}, P. Camus\inst{1},
    C. Giordano\inst{4}, R. Guesten\inst{7}, C. Hoffmann\inst{1},
    S. Leclercq\inst{8}, P. Mauskopf\inst{6}, K. F. Schuster\inst{8}.
          }
         \institute{Institut N\'eel, CNRS \& Universit\'{e} Joseph Fourier, BP 166, 38042 Grenoble, France
         \and
         SRON, Netherlands Institute for Space Research, Postbus 800, 9700 AV Groningen, Holland
         \and
         SRON, Netherlands Institute for Space Research, Sorbonnelaan 2, 3584 CA Utrecht, Holland
         \and
         Dipartimento di Fisica, Universit\'a di Roma La Sapienza, p.le A. Moro 2, 00185 Roma, Italy
         \and
          Laboratoire d'Astrophysique, Observatoire de Grenoble, BP 53, 38041 Grenoble, France
         \and
         Cardiff School of Physics and Astronomy, Cardiff University, CF24 3AA, United Kingdom
         \and
         Max-Planck-Institut f\"ur Radioastronomie, Auf dem H\"ugel 69, 53121 Bonn, Germany
         \and
         Institut de RadioAstronomie Millim\'etrique, 300 rue de la Piscine, 38406 Saint Martin d'H\`eres, France
             }

    \authorrunning{Monfardini et. al.}

   \date{Accepted June 11, 2010}


  \abstract
   {Current generation millimeter wavelength detectors suffer from scaling limits imposed by complex cryogenic readout electronics.  These instruments typically employ multiplexing ratios well below a hundred. To achieve multiplexing ratios greater than a thousand, it is imperative to investigate technologies that intrinsically incorporate strong multiplexing.  One possible solution is the kinetic inductance detector (KID).  To assess the potential of this nascent technology, a prototype instrument optimized for the 2 mm atmospheric window was constructed.  Known as the N\'{e}el IRAM KID Array (NIKA), it has recently been tested at the Institute for Millimetric Radio Astronomy (IRAM) 30-meter telescope at Pico Veleta, Spain.}
   {There were four principle research objectives:  to determine the practicality of developing a giant array instrument based on KIDs, to measure current in-situ pixel sensitivities, to identify limiting noise sources, and to image both calibration and scientifically-relevant astronomical sources.  }
   {The detectors consisted of arrays of high-quality superconducting resonators electromagnetically coupled to a transmission line and operated at $\sim$100 mK.  The impedance of the resonators was modulated by incident radiation; two separate arrays were tested to evaluate the efficiency of two unique optical-coupling strategies.  The first array consisted of lumped element kinetic inductance detectors (LEKIDs), which have a fully planar design properly shaped to enable direct absorbtion.  The second array consisted of antenna-coupled KIDs with individual sapphire microlenses aligned with planar slot antennas.  Both detectors utilized a single transmission line along with suitable room-temperature digital electronics for continuous readout.}
   {NIKA was successfully tested in October 2009, performing in line with expectations.    The measurement resulted in the imaging of a number of sources, including planets, quasars, and galaxies.  The images for Mars, radio star MWC349, quasar 3C345, and galaxy M87 are presented.  From these results, the optical NEP was calculated to be around $1 \times 10^{-15}$ W$/$Hz$^{1/2}$.  A factor of 10 improvement is expected to be readily feasible by improvements in the detector materials and reduction of performance-degrading spurious radiation.
}
   {}


   \maketitle
%

\section{Introduction}

Millimeter and sub-millimeter wavelength observations are at the forefront of cosmology and astrophysics.  They have proven to be an indispensable tool for understanding the early stages of star and galaxy formation.  Furthermore, continuum measurements provide a direct probe of the anisotropy of the cosmic microwave background, the Sunyaev-Zel'dovich effect in clusters of galaxies and the dust emission in the Galaxy and external galaxies.  Low-temperature bolometers have historically been utilized as detectors for these measurements. For this reason, bolometers have enjoyed decades of technical improvements and single pixels are now reaching fundamental limits.  Once intrinsic device limits have been achieved, greater sensitivity can only be realized by increasing the focal plane area and pixel count.

Current bolometer arrays such as MAMBO, LABOCA, SABOCA, APEX-SZ, SPT, SCUBA-2, Herschel PACS, and SPIRE now typically employ up to a few thousand pixels.  This scaling has resulted in increasingly complex readout electronics that must implement some form of multiplexing to reduce cryogenic wire-counts.  For example, time-domain multiplexing, or rapid switching, can be achieved using quantum point contact high-mobility transistors (\cite{Benoit:940}) or MOSFET (\cite{Reveret:32}).  In frequency-domain multiplexing, instead of rapid switching between separate wires, the bandwidth of a single wire is subdivided between multiple pixels.  This technique has been previously demonstrated by integrating superconducing quantum interference devices (SQUIDs) with a type of bolometer known as the transition edge sensor (TES) (\cite{yoon:371}, \cite{irwin:2107}, \cite{irwin:63}).  However, to realize the very large multiplexing ratios required for many-kilopixel arrays, it is necessary to develop detectors intrinsically adapted to strong frequency-domain multiplexing.  One potential device that achieves high sensitivity and is fundamentally compatible with frequency-domain multiplexing is the kinetic inductance detector (KID).

Based on high-quality, low-volume superconducting microwave resonators, KIDs were first proposed less than ten years ago for millimeter radiation detection (\cite{Day2003}).  Benefiting from intense research in quantum computing and fundamental physics (\cite{Hofheinz2009}; \cite{Grolosvsky6462}), development proceeded quickly and resulted in the first telescope measurement employing KIDs at the Caltech Submillimeter Observatory (\cite{Schlaerth2008}).  In view of a future large instrument and in order to test the feasibility of a receiver array based on KIDs, we have recently constructed a prototype instrument, known as the N\'{e}el IRAM KID Array (NIKA).  NIKA was recently tested at the 30-meter Institute for Millimeteric Radio Astronomy (IRAM) telescope at Pico Veleta, Spain.  Two complementary technologies were tested during this run: a 30-pixel lumped element kinetic inductance detector (LEKID) array (\cite{doyle:156}) and a 42 pixel antenna-coupled KID array.  Here we present an overview of the operating principles of the detector arrays, a summary of the principal instrument design elements, and the results from the first measurement run.


\section{Kinetic Inductance Detectors}

Kinetic Inductance Detectors have recently emerged as a promising alternative to classical bolometers for mm and sub-mm astronomy. In a bolometer, incident radiation is first converted into thermal phonons.  This increase in temperature is then transduced by a suitable thermistor, such as a doped semiconductor, TES, or magnetic calorimeter, into a measurable electrical signal.  However, to maintain a sufficiently long-lived thermal nonequilibrium between the bolometric element and the bulk substrate, complicated micromachined structures are required to reduce the thermal conductivity between these two systems.  This has necessitated the development of, for example, thin, large-area SiN membranes and spider-web geometries (\cite{gildemeister:868}).

\begin{figure}[h]
  \centering
   \includegraphics[angle=90, width=.95\linewidth]{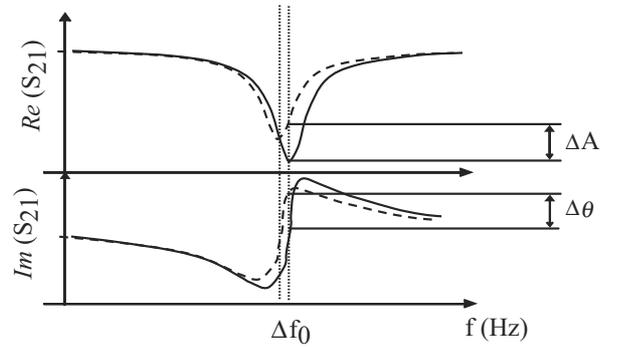}
      \caption{KID working principle. S$_{21}$ is the complex transmission and f$_0$ is the resonant frequency dependent upon the kinetic inductance, geometric inductance and capacitance.  The solid curves are the measured resonance under dark conditions and the dashed curves are under illumination. Phase ($\theta$) and amplitude (A) are measured simultaneously. }
         \label{Fig1}
\end{figure}

In a superconductor however, incident photons exceeding the gap energy result in pair breaking and a concurrent rise in quasiparticle density.  At sufficiently low temperature (T $\ll$ T$_c$) and in high quality films, these nonequilibrium quasiparticles have a long lifetime due to the low quasiparticle-phonon interaction.  For example, typical lifetimes in high quality aluminum films at 100 mK are $\sim$ 100 $\mu$s.  For certain specific geometries, the elevated quasiparticle density results in a change in the surface reactance of the material.  This is known as the kinetic inductance effect (\cite {Tinkham:1975}; \cite{Schmidt:1997}).  KIDs harness this changing reactance of the superconductor by embedding it in a high quality resonant circuit eletromagnetically coupled to a transmission line. Slight deviations in the kinetic inductance results in a shift in the resonant frequency of the device.  A measurement of the complex response of a KID with and without illumination is shown in \figref{Fig1}.  From this, the basic measurement procedure is readily understood.  First, a calibration sweep is taken to determine the shape of the resonance. Then a continuous wave drives the transmission line at the resonance frequency and the transmitted phase and amplitude of the wave is continuously monitored.  Shifts in the resonant frequency due to changes in illumination result in measurable changes in the transmitted amplitude and phase.

Compared with bolometers, KIDs have a number of distinct advantages.  First, as the quasiparticle population is naturally decoupled from the bulk insulating substrate at low temperatures, it is unnecessary to micromachine complicated support structures, greatly reducing fabrication complexity.  Second, in contrast with TES bolomters which operate at T $\sim$ T$_c$ where the sensitivity to temperature changes is greatest, KIDs operate at T $\ll$ T$_c$.  This leads to a significantly decreased sensitivity to temperature fluctuations and microphonic noise. Another advantage when compared to multiplexing techniques employing SQUIDs is that KIDs are significantly less sensitive to magnetic field fluctuations.  Finally, the resonant frequencies of the individual resonators can be easily controlled geometrically during array design.  For example, for $\lambda/4$ resonators, it is merely necessary to slightly adjust the resonator lengths during the array design to yield the desired frequency separation.  Along with the sharpness of the resonances due to the high-Q nature of the circuit, this allows many resonators to be packed into a limited bandwidth.  The result is that a very large number of pixels, potentially thousands, can be multiplexed on a single transmission line.  One caveat however, is that while the cryogenic wire-count is significantly reduced, it necessitates a sophisticated digital readout at room temperatures to both generate the necessary measurement tones and to continuously monitor the complex transmission.  Fortunately, recent advances in programmable electronics readily provide a solution to this impediment.

While KIDs clearly offer significant advantages over more canonical detection technologies, a number of technical challenges remain.  The choice of superconducting materials and substrates directly affect the cutoff frequency, sensitivity and intrinsic device noise.  For example, the cutoff photon frequency for detection, $\nu_{min}$, is determined by the superconducting gap:  $\nu_{min}={2\Delta_g}\,/\,h={3.528kT_c}\,/\,h$ according to Bardeen-Cooper-Schriefer (BCS) superconductivity theory at T $\ll$ T$_c$ (\cite{Tinkham:1975}).  Obvious options for mm and sub-mm wave detection are aluminium (T$_c=1.2$ K, $\nu_{min}=115$ GHz), titanium (T$_c=0.5$ K, $\nu_{min}=48$ GHz) and TiN$_x$ (T$_c=.8$ to 4.5 K, $\nu_{min} =$77 to 430 GHz) (\cite{Leduc:2010}). Real cutoffs can differ slightly due to material defects and deviations from BCS theory.  For constant energy photons, a reduced superconducting gap also results in a greater response per absorbed photon.  However, this comes at a cost of requiring lower operating temperatures.  The sensitivity of the array also increases with quasiparticle lifetime.  This last parameter is not only material dependent but process dependent as well, resulting in strict processing control requirements during fabrication.  Finally, the intrinsic noise has been found to be material dependent and most likely due to surface defects and oxide formation (\cite{Cooper:2004}; \cite{gao:102507}).

\begin{figure}[h]
   \centering
    \includegraphics[width=.95\linewidth]{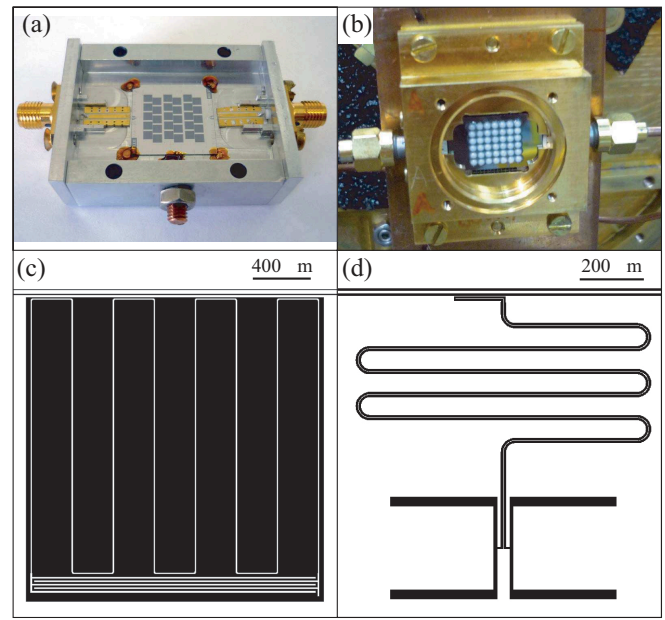}
      \caption{(Color online) Array and single-pixel images. (a) LEKID 30 pixel array and (b) antenna-coupled 42 pixel array with sapphire microlenses.  The dimensions of the focal plane are in both cases $\sim$ 1 cm$^2$, corresponding to a 45 arcsec field-of-view. (c) Single LEKID resonator design and (d) antenna-coupled resonator design.  For both designs, a high quality metallic film is first deposited onto a dielectric wafer.  A subsequent etching process is then used to remove metal in the black regions.}
         \label{Fig2}
\end{figure}

In order to leverage the high sensitivity of KIDs to pair-breaking radiation, it is necessary to first efficiently couple incident radiation into the superconducting resonator.  During the latest run, NIKA tested two distinct technologies for achieving this coupling:
\begin{enumerate}{}{}
\item LEKIDs. In this design, the resonator is composed of two sections: a long meander line that is primarily inductive and a capacitive section useful for tuning the resonant frequency.  Due to this unique geometry, the current in the meander section is nearly constant resulting in a position independent response to photon absorption.  Further, the geometry of the meander section allows it to appear as a solid absorber at the targeted wavelengths.  This design is fully planar and merely requires a backshort cavity situated at a calculated distance from the resonators.  For the current run, array fabrication consisted of a single 40 nm aluminum evaporation on a standard 270 $\mu$m thick low-resistivity silicon wafer cleaned using argon milling.  The film was then subsequently patterned using standard UV lithography and wet etching.  Two resonators were damaged during processing resulting in a pixel yield of 93\%.
\item Antenna coupled KIDs. A distributed, $\lambda / 4$ resonator is terminated with a slot antenna. The antenna concentrates the incident radiation into the high current-density region of the resonator.  This approach requires a microlens per pixel for efficient coupling.  A particular advantage to this design is the compatibility with frequency selective applications.  In this case, a 100\% pixel yield was achieved.
\end{enumerate}
The two NIKA arrays tested on the sky are displayed in \figref{Fig2}. The 42 pixel antenna-coupled array is designed to Nyquist sample the diffraction pattern at 2 mm wavelength ($0.5 \cdot F \lambda $).  However, while the pixel spacing is 1.6 mm required for Nyquist sampling, the lens diameter is 1.5 mm thus giving an array filling factor of 69\%.  For the 30 pixel LEKID array, in order to keep about the same 45 arcsec projected field-of-view, the final aperture ratio is approximately $0.65 \cdot F \lambda $.  In this case, the filling factor is 75\%.

\section{The NIKA instrument}

The NIKA instrument was custom designed for the IRAM 30-meter Nasmyth-focus telescope.  It utilizes a compact dilution cryostat with a cooling power of $\sim$ 100 $\mu$W at 100 mK.  This cryostat was engineered to meet both physical space requirements and to easily mate with the existing cabin optics.  Operation of the cryostat, including the gas handling system, is automated and is controlled remotely.  A rotating tertiary mirror (M3) allows the instrument to remain stationary during operation.

\begin{figure}[h]
    \centering
    \includegraphics[width=.95\linewidth]{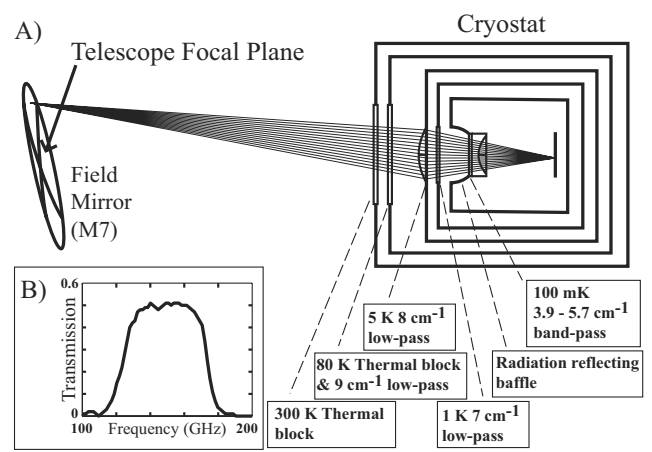}
    \caption{NIKA optical configuration. (a) The IRAM telescope focal plane coincides with the M7 tangential plane. M7 is a bi-polynomial mirror interfacing the NIKA instrument with the telescope optics.  (b) Calculated transmission of the filter stack, based on  measurements of the individual elements.  The bandpass is from 125 to 170 GHz.}
     \label{Fig3}
\end{figure}

The NIKA optical design is displayed in \figref{Fig3}.  The telescope focal plane is re-imaged with a bi-polynomial field mirror and two high density polyethylene (HDPE, n=1.56) lenses.  In order to reduce the aberrations introduced by the field mirror, the deflection angle is kept as low as possible, the lower limit being determined by the cryostat physical dimensions. The de-magnifying factor is $\sim$6, giving an effective aperture on the detectors plane of f/1.6.  In its present configuration, the instrument admits up to a 100 mm diameter focal plane with an aperture as low as f/1.6.  It is is thus able to accommodate the full 4.5 arcmin dish field-of-view.  The re-imaging optics are telecentric in image space, with the chief rays parallel to the optical axis.  Thus, the illumination on the focal plane is everywhere perpendicular to the detectors with the light cones subtending the same angle at every point.  This homogeneity eliminates any position dependence across the array arising from the overlap of the fixed angular-radiation pattern of the pixels and the incident radiation.

In order to eliminate any spurious off-axis radiation which can significantly degrade the detector performance, a novel cold baffling system was designed and fabricated.  It is based on a three-stage stack of properly shaped aluminum reflecting surfaces.  Each stage consists of a cross section of an ellipsoid shaped in such a way that any stray rays entering the optical system are back-reflected (\cite{benoit:702009}). In addition to this baffling system, infrared-blocking and band-defining filters were installed on the radiation screens at 80 K, 5 K, 1 K and 100 mK. These consist of multi-layer cross-mesh filters, each augmented with a layer of ZITEX\textsuperscript{\textregistered} film.  The bandpass of this filter stack is 125-170 GHz.

\begin{figure}[h]
    \centering
    \includegraphics[width=.95\linewidth]{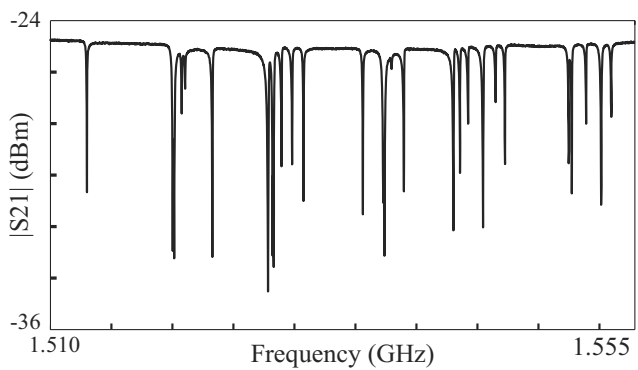}
      \caption{Feedline transmission (S21) frequency scan showing the resonances. Every dip corresponds to a single pixel.}
         \label{Fig4}
\end{figure}

\begin{figure}[h]
    \centering
    \includegraphics[width=.95\linewidth]{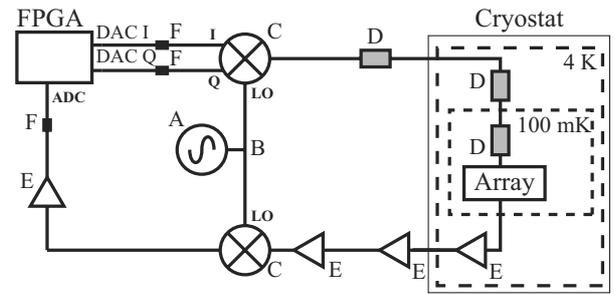}
      \caption{Basic measurement schematic.  The component labels correspond to: A) .01-8 GHz High-frequency synthesizer, B) Splitter, C) Mixer, D) Attenuator, E) Amplifier, and F) Low-pass filter.  Not shown is an acquisition computer connected to the FPGA via a communication link.}
         \label{FigElectric}
\end{figure}

An electrical transmission measurement of the LEKID array is shown in \figref{Fig4} and a general circuit schematic used for frequency-multiplexed array readout is displayed in \figref{FigElectric}.  The readout operating principles have been previously described in detail (\cite{yates:042504}, \cite{swenson:84}), however it is useful to describe the basic procedure here.  A field-programmable gate array (FPGA) is used to generate a waveform consisting of a sum of sinusoidal waves; the frequency spacing of the individual tones corresponds to the frequency spacing of the resonator array.  This waveform and a 90$^{\circ}$ phase-shifted copy is fed to two high-resolution digital-to-analog converters (DACs).  After low-pass filtering, these signals are then used to drive the in-phase (I) and quadrature (Q) ports of an up-converting IQ mixer.  The mixer local oscillator (LO) port is simultaneously powered by a low-noise, high-frequency synthesizer.  An IQ mixer is employed to suppress the parasitic lower-sideband and carrier frequencies.  The mixer output is then fed into the cryostat where low-temperature attenuators reduce the thermal Nyquist noise from the room-temperature electronics.  Superconducting NbTi coaxial cabling is used at the lowest temperature stages to reduce thermal loading on the coldest stage ($^3$He/$^4$He mixing chamber).  The signal then interacts with the resonator array where the response of the individual resonators modulates the waveform.  The resulting waveform is amplified by a low-noise amplifier (T$_{noise}$ = 3-5 K) mounted at a physical temperature of 4.5 K.  The signal then exits the cryostat and is down-converted by a second mixer driven by the same high-frequency synthesizer used for up-conversion.  The output of the mixer is then properly scaled, low-pass filtered and fed into an analog-to-digital converter (ADC).  The individual tones are then digitally separated.  The resulting in phase and quadrature components of the individual sinusoids, each pair corresponding to the response of a single pixel, are returned over a communication link to a control computer at a rate of 5-25 Hz.


An important measurement consideration is the sensitivity of the resonances to background radiation and the local magnetic field.  The optical design was optimized to reject off-axis and out-of-band radiation.  Further, a single layer of magnetic shielding was wrapped around the 1 K radiation screen to reduce magnetic field effects.  Despite these precautions, the resonance frequencies of both the LEKID array and the antenna-coupled array were weakly sensitive to the telescope position.  This most likely resulted from the motion of trapped flux in the superconducting resonators.  It was thus necessary to recalibrate the array after making a large positional shift to target a new source.  During the first NIKA test run, a semi-automatic calibration procedure was implemented to recenter the measurement electronics on the array resonance frequencies. For the future, a fast, fully-automated calibration procedure will be used to ensure proper array operation.  Also, a new filter stack and magnetic shield are currently being designed to further reduce these parasitic effects.


\section{Results and perspectives}

The first NIKA engineering run took place in October 2009. The instrument was installed in the receiver cabin of the IRAM 30-meter telescope at Pico Veleta, Spain and operated remotely from the control room.  The cool down of the instrument was also performed remotely, taking approximately 18 hours to reach the operating temperature of 100 mK.  The observations were carried out during daytime and were divided into two separate slots.  The first slot consisted of two days of measurement with the antenna-coupled KID array.  A non-observing day was then used to warm the cryostat, mount the LEKID array and re-cool the system.  This second array was then measured for the remaining four days.  The primary observing mode for both arrays was on-the-fly scanning while tracking a target source.  The elevation steps between sweep lines was typically 6 arcsec; the measured beam FWHM is 19 arcsec.  A second observing mode chopped the secondary mirror (M2, D = 2 m) to alternate between the source and a dark position on the sky at a rate of about 1 Hz.

\begin{figure}[h]
   \centering
    \includegraphics[width=.95\linewidth]{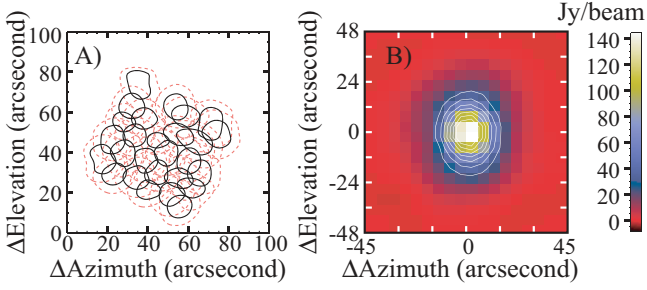}
      \caption{(Color online) Array calibration.  (a) LEKID beam pattern used for pixel identification.  The contours shown are at 80\% (black, solid) and 50\% (red, dashed) of maximum flux.  The sources are slightly elongated in the cross-scan direction due to mild in-scan data filtering used to remove 1/f noise.  (b) Primary calibration source.  Coadded map of Mars using 20 maps taken with the LEKID array.  The contours range linearly from 20\% to 80\% of the maximum flux.
              }
         \label{MarsFigure}
\end{figure}

\begin{figure}[h]
   \centering
    \includegraphics[width=.95\linewidth]{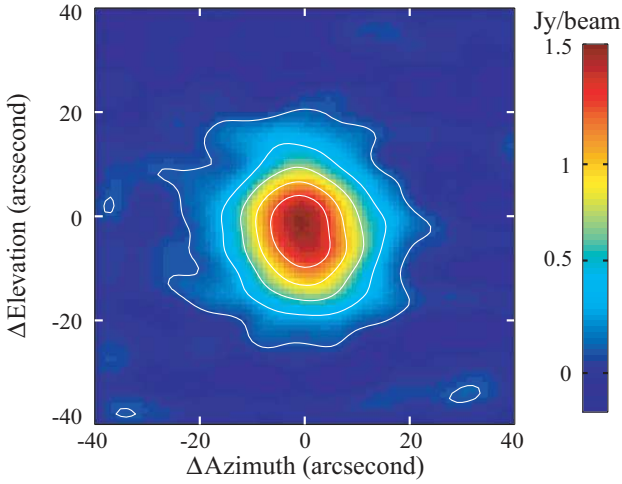}
      \caption{(Color online) Coadded map of radio star MWC349 used as a secondary calibrator.  This image was taken using the antenna-coupled array.
              }
         \label{2ndaryCalibrator}
\end{figure}

A primary objective for this run was to image both calibration and scientifically-relevant astronomical sources.  For both detection arrays, the list of target sources included standard calibrators (e.g. planets) and a selection of fainter sources (e.g. AGNs, radio stars, and galaxies). The main calibrator was Mars (7.47 arcsec angular diameter, 166 Jy) which, when diluted in the 19 arcsec measured beam, represents an equivalent temperature of 21 K. It is seen with an average S/N of 440 Hz$^{-1/2}$ per pixel.  An image of Mars can be seen in \figref{MarsFigure}.  Using pure phase readout, the linearity of the calibration was established with a secondary 1.55 Jy calibrator MWC349 shown in \figref{2ndaryCalibrator} to be within a few percent.

\begin{figure}[h]
   \centering
    \includegraphics[width=.95\linewidth]{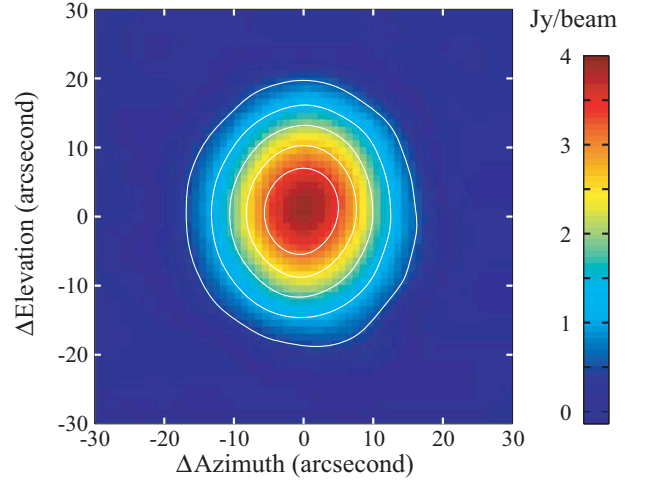}
      \caption{(Color online) Coadded map of quasar 3C345 with a measured flux of 3.9 Jy on October 22th at 15.25 UT. This image was taken using the antenna-coupled KID array.
              }
         \label{3C345}
\end{figure}

\begin{figure}[h]
   \centering
    \includegraphics[width=.95\linewidth]{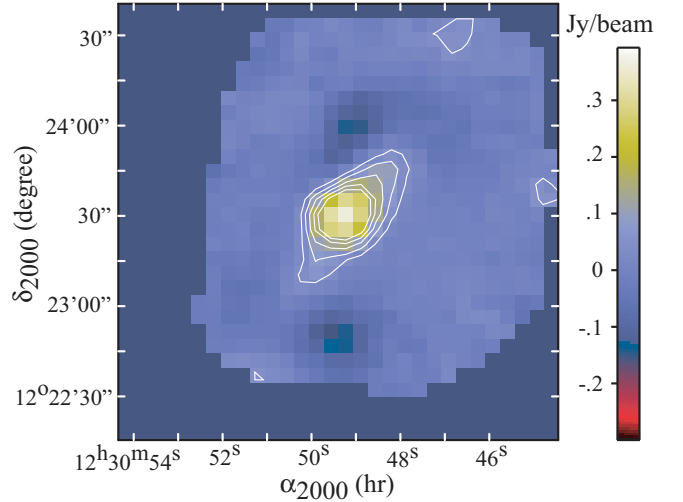}
      \caption{(Color online) Coadded map of galaxy M87 using 19 individual KID maps. This radio source
clearly shows extensions. The two symmetrical negative lobes are due to temporal low-pass filtering during processing. This image was taken using the LEKID array.
              }
         \label{M87}
\end{figure}

Using this calibration, we observed a number of sources.  An image of the point-like quasar 3C345 is shown in \figref{3C345}.  An image of galaxy M87, demonstrating extended-source mapping capabilities, is shown in \figref{M87}.  This image was obtained by integrating for 0.5 hours.  The noise was measured to be 6 mJy$/$Beam. The point source at the center has a flux of 310$\pm$4 mJy.  As expected, the source is extended.  We targeted the gamma ray burst GRB091024 on October 24 at16.33 UT (7.40 hours after the burst), and integrated for 1.5 hours under average sky conditions.  A mJy-scale, 150 GHz photometric upper limit obtained a few hours after the burst is potentially useful in constraining the fireball parameters for GRB091024 (z=1.09).   The GRB was not detected with an upper limit of around 24 mJy (3$\sigma$).  The central observation frequency was 148 GHz.

\begin{figure}[h]
   \centering
    \includegraphics[width=.95\linewidth]{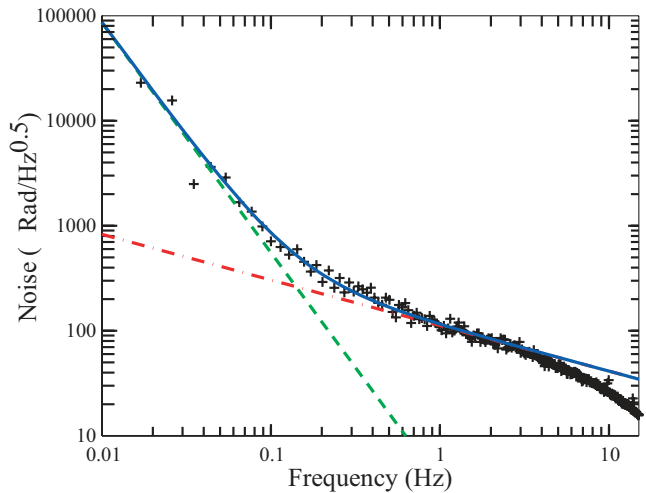}
      \caption{(Color online) LEKID phase-noise spectrum measured during a total power scan (+, black) under normal daytime conditions ($\tau \sim$.2 at 225 GHz).   The roll-off above 5 Hz is related to the readout electronics rate. The solid line (blue) is a fit of $N_{\theta} = A f^\alpha + B f^\beta$ to the data below 5 Hz ($\alpha = -0.43$, $\beta = -2.19$).  The dashed line (green) is a plot of the fitted $A f^\alpha$ term which is predominately caused by sky fluctuations. The dash-dot line (red) is a plot of the $B f^\beta$ term which is dominated by intrinsic detector noise.  No change in the intrinsic detector noise or spurious electrical noise was observed upon moving from the laboratory to the telescope environment.
              }
         \label{Fig5}
\end{figure}

In the \figref{Fig5}, we present the single-pixel phase noise measured during a typical total-power scan. The intrinsic detector phase-noise was determined to be $N_{\theta} \propto f^{-0.43}$ which is compatible with a typical resonator noise spectrum $N_{\theta} \propto f^{-0.5}$ that has previously been observed (\cite{gao:102507}).  This spectral slope has been explained with random variations of the effective dielectric constant due to the presence of quantum two-level systems (TLS) surrounding the resonators (\cite{Cooper:2004}). These TLS are most likely located at the interface between the metallic film and the substrate; on the substrate surface; and in the thin AlO$_x$ layer present on the aluminum film (\cite{gao:152505}). The low-frequency sky noise is fitted to $N_{\theta} \propto f^{-2.19}$.  The noise equivalent power (NEP) of the detector was determined from the measured response to Mars and the corresponding noise spectrum.  It was determined to be about $1.2 \times 10^{-15}$W$/$Hz$^{1/2}$ at 1 Hz in a single polarization. The response of the detectors is frequency independent for modulated signals in the bandwidth of interest for long-wavelength astronomy.  Thus, the frequency dependence of the NEP is directly proportional to the detector noise shown in \figref{Fig5}.   The noise equivalent temperature (NET) of the best array was determined to be $\sim$15 mK$\cdot$s$^{1/2}$ per beam.  This corresponds to a detector noise equivalent flux density (NEFD) of 190 mJy$\cdot$s$^{1/2}$ or an NEFD of 120 mJy$\cdot$s$^{1/2}$ per beam.  Note that these values should be divided by two for comparison with polarization insensitive devices.  This has been calculated directly from the summed map RMS and hence is a direct estimation of the full system NEFD at the frequencies of interest for a single-polarization, total-power scan.  Further, the dynamic range of the KIDs was sufficient for normal operation.  For example, under normal daytime observation ($\tau \sim$.2 at 225 GHz), a full skydip (1.1 to 1.8 air-masses) was performed without pixel saturation.

During this first engineering run, the NIKA instrument performed in line with preliminary laboratory measurements and met the first-stage design goals.  The detector arrays used during this test were comparable in size and sensitivity.  The antenna-coupled array admits the possibility of a frequency-selective operation while the LEKID array does not require individual lenses and hence should scale more easily.  The choice in detector technology thus is application specific and dependent upon future improvements in the intrinsic device noise.  The sensitivity during this measurement was a factor of 5 to 10 above the best ground-based arrays currently in use. For example, the MAMBO-2 receiver has 117 bolometers with 35 mJy$\cdot$s$^{1/2}$ sensitivity at 250 GHz. We aim at reaching at least this sensitivity while covering a 5 to 8 arc-min diameter field-of-view available before vignetting.  That would allow a minimum factor of 3 to 10 increase in mapping speed and better handling of confusion and sky noise removal. To achieve this performance, arrays of thousands of pixels must be assembled.  The sensitivity goal is expected to be achievable in the near future by increasing the system optical efficiency using anti-reflective coatings on the lenses and windows; reducing spurious radiation with a redesigned baffling structure; reducing magnetic field fluctuations by increasing the magnetic shielding; reducing the detector phase-noise by eliminating contaminant oxide layers and reducing local electric fields; increasing the pixel responsivity by using higher quality metallic films for the resonators and finally optimizing the detector optical coupling.  A second NIKA engineering run is planned for 2010, which will apply the aforementioned system improvements and utilize significantly larger detector arrays.

\begin{acknowledgements}
      Thanks to Santiago Navarro, Clemens Thum, Jean-Yves Chenu, Carsten Kramer, Walter Brunswig, Javier Lobato and all the IRAM staff for the excellent technical support during the run.
      Part of this work was supported by grant ANR-09-JCJC-0021-01 of the French National Research Agency, the Nanosciences Foundation of Grenoble and R\'egion Rh\^one-Alpes (program CIBLE 2009).
\end{acknowledgements}

\end{document}